# Superfluid-like TO Responses in Rotating Solid Helium

J. Choi[1], T. Tsuiki[2,3], D. Takahashi[4,3], H. Choi[1], K. Kono[3], K. Shirahama[2,3] and E. Kim[1*]

[1] Department of Physics and Center for Supersolid and Quantum Matter Research, Korea Advanced Institute of Science and Technology (KAIST), 291 Daehak-ro, Yuseong-gu, Daejeon 34141, Republic of Korea
[2] Department of Physics, Keio University, 3-14-1 Hiyoshi, Yokohama-shi, Kanagawa 223-8522, Japan
[3] Center for Emergent Matter Science, RIKEN, 2-1 Hirosawa, Wako, Saitama 351-0198, Japan
[4] Center for Liberal Arts and Sciences, Ashikaga Institute of Technology, 268-1 Omae, Ashikaga, Tochigi 326-8558, Japan

**The irrotational nature of superfluid helium was discovered through its decoupling from the container under rotation. Similarly, the resonant period drop of a torsional oscillator (TO) containing solid helium was first interpreted as the decoupling of solid from the TO and appearance of supersolid[1-3]. However, the resonant period can be changed by mechanisms other than supersolid, such as the elastic stiffening of solid helium[4,5] that is widely accepted as the reason for the TO response. To demonstrate the irrotational nature more directly, the previous experiments superimposed the dc rotation onto the TO and revealed strong suppression on the TO response without affecting the shear modulus[6,7]. This result is inconsistent with the simple temperature-dependent elasticity model and supports the supersolid scenario. Here, we re-examine the rotational effect on solid helium with a two-frequency rigid TO to clarify the conflicting observations. Surprisingly, most of the result of previous rotation experiments were not reproduced. Instead, we found a very interesting superfluid-like irrotational response that cannot be explained by elastic models.**

We constructed a two-frequency rigid TO to validate previous DC rotation experiments as shown in Figure 1. If the TO is not sufficiently rigid, the TO period change due to the elastic stiffening of the solid helium is enormously amplified[8-11]. The period drop reported in the previous rotation experiment was unusually large when compared to other rigid TOs. We note that a non-rigid TO is easily affaced by the vibration noise generated during dc rotation and may show a rotation effect. Besides, the rigid TO can be operated at two resonance frequencies, which allows the frequency-dependence analysis on the TO response. The analysis can distinguish the contribution of elastic stiffening in the period change from that of superfluid. For example, the "overshoot effect" of the linear elastic effect gives a period drop proportional to the square of the TO frequency, while the superfluid-induced period drop does not change with frequency. The two-frequency rigid TO used in this experiment reported a frequency-independent non-classical rotational inertia (NCRI) fraction of about $4 \times 10^{-6}$[12] that does not exceed the previously proposed supersolid upper limit[13,14]. We measured the period and amplitude of the TO during increasing temperature from 20 mK to 500 mK while maintain a uniform rotational speed $\Omega$. The TO was attached to the mixing chamber of the rotating dilution refrigerator at RIKEN to superimpose the DC rotation. Polycrystalline solid helium samples were grown using the block-capillary method.

Figure 2 shows the temperature dependence of the period drop $dP$ divided by mass loading /$\Delta P(T)$ of solid helium, $dP/\Delta P(T)$, and dissipation $Q^{-1}(T)$ in both in-phase and out-of-phase modes. At $\Omega = 0$, the typical characteristics of a rigid TO were successfully reproduced. In the in-phase (out-of-phase) mode, the reduction of the TO period deigns at about 250 mK (325 mK) and saturates at approximately 50 mK (60 mK). The frequency dependence of the onset temperature is consistent with that in the other experiments using a double-frequency TO[15] and PZT[7]. In the in-phase and out-of-phase modes, the TO resonant period decreased by 1.10 ns and 0.87 ns, respectively. After subtracting the background signal of the empty cell, the $dP/\Delta P(T)$ values of the two modes at 25 mK are $dP_{in}/\Delta P_{in} = 2.86 \times 10^{-5}$ and $dP_{out}/\Delta P_{out} = 1.48 \times 10^{-4}$, where the subscripts indicate the aspecific modes. Dissipation $Q^{-1}(T)$ shows local maxima where the period of TO changes most steeply. The maximum size of dissipation $Q_{in}^{-1}$ ($Q_{out}^{-1}$) is approximately $1.43 \times 10^{-7}$ ($1.93 \times 10^{-7}$), which was much smaller than that of typical TOs[12-14].

We also examined the effect of DC rotation on the TO response by systematically increasing the rotational speed. To concentrate on the DC rotation effect, the amplitude of AC torsional oscillation was fixed to the minimum value. As shown in Figure 2, the TO response measured in (a) the in-phase and (b) out-of-phase mode shows a striking contrast. In the in-phase mode, the period drop was clearly suppressed as the dc rotation speed increased, but no such rotation effect was found in the out-of-phase mode. At 4 rad/s, the magnitude of $dP_{in}/\Delta P_{in}$ at 30mK decreases to $2.53 \times 10^{-5}$, which is about 12% smaller than the value without rotation. On the other hand, the temperature curves of $dP_{out}/\Delta P_{out}$ measured at various rotational speeds coincide well. The difference in $dP_{out}/\Delta P_{out}$ at 30 mK measured at 0 and 4 rad/s rotation is smaller than $2 \times 10^{-6}$, which is only 1.5% of the original value. Figure 2-(c) and 2-(d) show the temperature-dependent dissipations measured in two modes, respectively. The dissipation measured under various DC rotational speeds seems to be identical. No significant enhancement of the dissipation reported in previous rotation experiments was observed.

Although the suppression of the period shift in the in-phase mode is similar to the previous rotation experiment, the absence of the suppression in the out-of-phase mode shows clear distinction. This results indicates that the rotation effect cannot be caused by the shear modulus stiffening since the elastic effect will be greater in the higher frequency mode. No rotation effect in the dissipation also disproves the elastic models because the change in rotational susceptibility associated with elasticity must affect both period and dissipation[16,17].

On the other hand, a possible clue to understand previous rotation experiments can be found in the rigidity of TO. The previous non-rigid TO was very sensitive to the shear modulus changes in solid helium because the solid helium in the TO cell links the inner and outer structures[8]. If the vibration noise due to the rotation increases the shear stress in the tangential direction $\sigma_t$, the non-rigid TO response may mimic the so-called rotation effect. As the rotation speed increases, the vibration increases accordingly, resulting in greater suppression of the TO response. However, $\sigma_t$ from the rotation noise and the stress in the radial direction $\sigma_r$ due to the centrifugal force are few orders of magnitude smaller than the critical stress required to cause elastic softening[7]. This is consistent with the lack of the rotation effect in previous shear modulus measurements[6]. Therefore, it is necessary to find a new alternative explanation that can justify both the lack of the rotation effect in the shear modulus measurements and the significant suppression in the TO response. We assume the complicated structure of the previous TO may amplify the vibration noise effect and lead to the sizable dc rotation effect.

In the previous rotation velocity-sweep experiment, the step-like suppression of the TO period was found along with the hysteresis between the rotation speed decrease and consequent increase. This is interpreted as evidence of possible quantized vortices and Meissner state[7]. To verify this, the rotation-sweep experiment was performed again. The results are shown in Figure 3-(a). Neither staircase-like suppression nor hysteresis were found in both modes. In addition, we observed a small but reproducible jump of the TO response at around 2 rad/s where the most pronounced rotation effect was found in previous experiments. This supports the TO used in previous experiments was very sensitive to the rotational noise that could have been amplified at the corresponding rotation speed.

To better understand the suppression of the DC rotation response in the in-phase mode, we investigated the frequency dependence in various dc rotational speeds. Since the rigid double-torus TO removes all complicated coupling between viscoelastic solid and the TO structure, only tow contributions are considered for quantitative analysis: (a) the frequency independent contribution by the superfluid component $(dP/\Delta P)^{ind}$ and (b) the frequency dependent term $(dP/\Delta P)^{dep} \sim f^2$ due to the elastic "overshoot" effect[8]. Figure 4 shows (a) $(dP_{in}/\Delta P_{in})^{ind}$ and $(dP_{in}/\Delta P_{in})^{dep}$ measured at the lowest ac rim velocity under various rotational speeds. At 30mK, the frequency-independent term was calculated to be about $5.01 \times 10^{-6}$, slightly larger than the range set in the current study using the same TO[12]. This increase may be associated with much higher $^3$He concentrations of this solid sample. It is particularly surprising that frequency-independent terms are gradually suppressed as the DC rotation speed increases. Under 4-rad/s rotation, the magnitude of the frequency independent term at 30mK decreases to $2.86 \times 10^{-6}$. On the other hand, the frequency dependent term remains unchanged at around $2.15(\pm 0.04) \times 10^{-5}$ until 4 rad/s.

To investigate whether the rotation effect is related to the ac rim velocity of TO, the same frequency analysis was performed on various ac rim velocities. The effect of constant dc rotation of 3 rad/s on the To response was compared with that of the stationary state. As shown in Figure 5-(a) and 5-(b), the TO response obtained essentially the same as that of the lowest rim velocity. The frequency independent term is unchanged even as ac rim velocity increases. On the other hand, it shows a clear suppression as the dc rotation speed increase. Under 3-rad/s rotation, $(dP_{in}/\Delta P_{in})^{ind}$ at four different ac rim velocities is about $1.86(\pm 0.37) \times 10^{-6}$ smaller than that in the stationary state. One the other hand, the frequency dependent term $(dP_{in}/\Delta P_{in})^{dep}$ is strongly suppressed as ac rim velocity increases without being affected by dc rotation. These experimental results demonstrate that the frequency dependent TO response is due to the elastic change of solid helium.

The period shift due to ac TO oscillation is explained by the oscillating shear stress that separates the $^3$He impurity from the dislocation and reduces the shear modulus[4,5]. In contrast, constant dc rotation does not cause shear stress on solid helium. Thus, it is surprising that the frequency independent term is not unchanged by torsional oscillation, but clearly reduced by uniform dc rotation. In addition, the suppression differs significantly from the conventional elastic effect in terms of critical temperature and velocity. The frequency dependent term $(dP_{in}/\Delta P_{in})^{dep}$ gradually changes from 300mK, while the frequency independent term $(dP_{in}/\Delta P_{in})^{ind}$ suddenly appears at 100mK. The threshold rotation speed for the suppression is also about two orders of magnitude greater compared with that of ac oscillation as shown in Figure 3-(b). Thus, the dc rotation effect on $(dP_{in}/\Delta P_{in})^{ind}$ is clearly distinguishable from the effect of ac oscillation due to the change in shear modulus.

The small but distinct frequency independent term $(dP_{in}/\Delta P_{in})^{ind}$ and its reduction by dc rotation were consistently observed in three solid samples with pressures of 32, 36 and 47 bar. These experimental results can only be explained by superfluidity, which is irreconcilable with simple elasticity models. Thus, we investigated how this superfluidty can exist in solid helum. The polycrystalline solid helium obtained by a blocked capillary method contains numerous topological defects such as dislocations or grain boundaries that can allow superflow[18]. Therefore, we investigate whether or not the presence of topological defects adequately explain the irrotational feature, the existence of $(dP_{in}/\Delta P_{in})^{ind}$ and its suppression under rotation.

First, if the dislocation of hcp solid helium can exhibit superfluidity, a network of randomly-interconnected one-dimensional dislocations may account for the rigid TO response. Both screw and edge dislocations are known to have superfluid cores[19,20]. The dislocation network may also be consider as an interconnected vortex tangles which may lead to the percolation of circulating supercurrent through dislocations with pinned vortex cores. This quasi-superfluid state, called the Shevchenko state[21], can be described by low viscosity fluid mimicking superfluid. Anderson also suggested that the superfluid flowing through the dislocation network or grain boundary can be understood as a vortex liquid model[22]. In order to examine the validity of this model, one can estimate the dislocation density $n_d$ required to explain the apparent superfluid fraction of $n_S^{3D}/n = n_d(n_S^{1D}/n) \sim 5 \times 10^{-6}$, where $n_S^{3D}$ and $n_S^{1D}$ is the 3D and 1D superfluid density, respectively. Applying the helium number density of $n \sim 0.0287$ per 1 Å$^3$ and the $n_S^{1D}$ value of 1 atom per 1 Å obtained from the grand-canonical Monte Carlo simulation[19], the required $n_d$ is about $1.4 \times 10^9 \ cm^{-2}$. The experimental measurements of $n_d$ for a single crystal ⁴He range from $3 \times 10^5/cm^{-2}$ [23] to $6 \times 10^9/cm^{-2}$ [24]. Thus, the quantitative estimation could support the superfluid dislocation network scenario. The drawback, however, is that the rotation influence on dissipation expected together with the suppression in the period shift has not been observed in our measurements.

Second, the grain boundary, another type of topological defects, can also be considered an array of tilted edge or screw dislocations. Path-integral Monte Carlo studies have found that the grain boundary of polycrystalline solid ⁴He becomes superfluid at a sufficiently low temperature around 0.5K[25]. Similar to other polycrystals, grain boundary premelting in hcp solid ⁴He was observed under high pressure condition above 1 kbar[26]. A thin premelted liquid film formed at the grain boundary can exhibit superfluidity through the two-dimensional BKT phase transition[27]. Since the amount of helium contained in TO is fixed, the apparent superfluid density $\rho_s = n_S^{3D}/n$, and the phase transition temperature $T_{BKT}$, depend on the size of $R_0$, the average characteristic diameter. Considering the spherical grain model[27], $R_0$ is given as $R_0 \cong (6m^2/\pi\rho\hbar^2)(k_B T_{BKT}/\rho_s)$, where $m$ is the atomic mass of ⁴He and the helium density $\rho$ is about 214 kg/m³. Using $\rho_s = 5 \times 10^{-6}$ and $T_{BKT} \sim 0.1K$, one can obtain $R_0$ to be a realistic value of about $10 \ \mu m$. Nevertheless, it is not consistent with the BKT transition model since there is no strong rounding effect to appear near the BKT phase transition temperature due to uneven film thickness and no additional rotation effect on the dissipation peaks.

Third, the most radical interpretation of the dc rotation induced suppression is the genuine bulk supersolid state. As described earlier, the superfluid density measured by the frequency independent term is very small, but it certainly exists. This minute size of superfluid can be screened by large shear modulus changes of solid helium at low temperatures, so that it is difficult to be detected precisely in a soft single-resonant-mode TO. Introduction of the rigid double-frequency TO allows us to distinguish the frequency independent term and striking rotation effects. Assuming that the supersolid state exists, $T_c \sim 90 mK$, $v_c \sim$ 1.5cm/s, and $\rho_S/\rho \sim 5 \times 10^{-6}$ for a 32-bar polycrystalline solid sample. However, this supersolid density is much higher than the upper limit set by the previous mass-flux experiments, $v_c \cdot (\rho_s/\rho) \sim 2.5 \times 10^{-8}$cm/s[28].

Finally, the Manchester group also performed a dc rotation experiment using a rigid TO, but unlike our experiments, no rotation effect was found[29]. Nevertheless, the dc rotation effect is not substantial at the low rotational speeds of about 2 rad/s, so that it can be masked by the elastic effects of about $10^{-4}$. The stability of the TO under rotation is also important for the clarification of the very small discrepancies in the response. In addition, we aware that the recent shear modulus experiments reported an astounding rotation effect; shear modulus was decreased by 15% at 4 rad/s[30]. At this moment, we do not understand why this incompatible rotation effect has been found. We believe that the shear modulus change of about 15% must have been detected in our TO experiments. However, it is very interesting that there is no rotation effect in their own Young's modulus measurement, which is consistent with our experimental results.

In conclusion, most of the results reported in previous rotation experiments[6,7] were not reproduced in a new experiment using rigid TO. This implies that the effect found in non-rigid TO can be explained by the change in shear modulus. Instead, we found a very intriguing rotation effect on the frequency-independent term, which is the unequivocal evidence of superfluidity. We cannot provide any convincing explanation that clearly understand all the experimental results. Further investigation is crucial to identify the fascinating anomaly.

## ACKNOWLEDGEMENTS

This work is supported by the National Research Foundation (NRF) of Korea grant funded by the Ministry of Science, ICT and Future Planning (MSIP) through the Creative Research Initiatives (2007-0054-848) and the Japan Society for the Promotion of Science (JSPS) through a Grant-In-Aid Science Research. J. C. greatly thank the POSCO TJ Park Foundation for financial support and generosity through the TJ Park Science Fellowship. J. C. also acknowledges the support of RIKEN IPA program during the collaboration. D. T. acknowledges the support from Takahashi Industrial and Economic Research Foundation.

## AUTHOR CONTRIBUTIONS

E. K. designed and managed the project. J. C. prepared the rigid torsional oscillator. J. C. and T. T. performed low temperature measurements, under on-site supervision of D. T., K. S. and K. K.. H. C. provided fruitful advices for data analysis. J. C. and E. K. wrote the manuscript. All authors discussed the result and manuscript preparation.

# REFERENCES

1. Boninsegni, M. & Prokof'ev, N. V. Supersolids: what and where are they?. *Rev. Mod. Phys.* **84**, 759-776 (2012).

2. Kim, E. & Chan, M. H. W. Probable observation of a supersolid helium phase. *Nature* **427**, 225-227 (2004).

3. Kim, E. & Chan, M. H. W. Observation of superflow in solid helium. *Science* **305**, 1941-1944 (2004).

4. Day, J. & Beamish, J. Low-temperature shear modulus in solid $^4$He and connection to supersolidity. *Nature* **450**, 853-856 (2007).

5. Syshcehko, O. Day, J. & Beamish, J. Frequency dependence and dissipation in the dynamics of solid helium. *Phys. Rev. Lett.* **108**, 105302 (2012).

6. Choi, H., Takahashi, D., Kono, K. & Kim, E. Evidence of supersolidity in rotating solid helium. *Science* **330**, 1512-1515 (2010).

7. Choi, H., Takahashi, D., Choi, W., Kono, K. & Kim, E. Staircaselike suppression of supersolidity under rotation. *Phys. Rev. Lett.* **108**, 105302 (2012).

8. Reppy, J. D., Mi, X., Justin, A. & Mueller, E. J. Interpreting torsional oscillator measurements: effect of shear modulus and supersolidity. *J. Low Temp. Phys.* **168**, 175-193 (2012).

9. Beamish, J. R., Fefferman, A. D., Haziot, A., Rojas, X. & Balibar, S. Elastic effect in torsional oscillators containing solid helium. *Phys. Rev. B* **85**, 180501 (2012).

10. Maris, H. J. Effect of elasticity on torsional oscillator experiments probing the possible supersolidity of helium. *Phys. Rev. B* **86**, 020502 (2012).

11. Shin, J., Choi, J., Shirahama, K. & Kim, E. Simultaneous investigation of shear modulus and torsional oscillation of solid $^4$He. *Phys. Rev. B* **93**, 214512 (2016).

12. Choi, J., Shin, J. & Kim, E. Frequency-dependent study of solid $^4$He contained in a rigid double-torus torsional oscillator. *Phys. Rev. B* **92**, 144505 (2015).

13. Kim, D. Y. & Chan, M. H. W. Upper limit of supersolidity in solid helium. *Phys. Rev. B* 90, 064503 (2014).

14. Eyal, A., Mi, X., Talanov, A. V. & Reppy, J. D. Search for supersolidity in solid 4He using multiple-mode torsional oscillators. *Proc. Natl. Acad. Sci. USA* **113**, 3203-3212 (2016).

15. Aoki, Y., Graves, J. C. & Kojima, H. Oscillation frequency dependence of nonclassical rotation inertia of solid 4He. *Phys. Rev. Lett.* **99**, 015301 (2007).

16. Hunt, B., Pratt, E., Gadagkar, V., Yamashita, M., Balatsky, A. V. and Davis, J. C. Evidence for a superglass state in solid $^4$He. *Science* **324**, 632-636 (2009).

17. Nussinov, Z., Balatsky, A. V., Graf, M. J. & Trugman, S. A. Origin of the decrease in the torsional-oscillator period of solid $^4$He. *Phys. Rev. B* **76**, 014530 (2007).

18. Pollet, M., Boninsegni, M., Kuklov, A. B., Prokof'ev, N. V., Svistunov, B. V. & Troyer, M. Local stress and superfluid properties of solid 4He. *Phys. Rev. Lett.* **101**, 097202 (2008).

19. Boninsegni, M., Kuklov, A. B., Pollet, L., Prokof'ev, N. V. & Svistunov, B. V. Luttinger liquid in the core of a screw dislocation in helium-4. *Phys. Rev. Lett*. **99**, 035301 (2007).

20. Soyler, S. G., Kuklov, A. B., Pollet, L., Prokof'ev, N. V. & Svistunov, B. V. Underlying mechanism for the giant isochoric compressibility of solid $^4$He. *Phys. Rev. Lett.* **103**, 175301 (2009).

21. Shevchenko, S. I. On one-dimensional superfluidity in bose crystals. *Fiz. Nizk. Temp.* **13**, 115-131 (1987).

22. Anderson, P. Two new vortex liquids. *Nature Physics* **3**, 160-162 (2007).

23. Iwasa, I., Araki, K. & Suzuki, H. Temperature and frequency dependence of the sound velocity in hcp $^4$He crystals. *J. Phys. Soc. Jpn.* **46**, 1119-1126 (1979).

24. Tsuruoka, F. & Hiki, Y. Ultrasound attenuation and dislocation damping in helium crystals. *Phys. Rev. B* **20**, 2702-2720 (1979).

25. Pollet, L., Boninsegni, M., Kuklov, A. B., Prokof'ev, N. V., Svistunov, B. V. & Troyer, M. Superfluidity of grain boundaries in solid $^4$He. *Phys. Rev. Lett.* **98**, 135301 (2007).

26. Franck, J. P., Korneisen, K. E. & Manuel, J. R. Wetting of fcc $^4$He grain boundaries by fluid $^4$He. *Phys. Rev. Lett.* **50**, 1463-1466 (1983).

27. Gaudio, S. Cappelluti, E. Rastelli, G. & Pietronero, L. Finite-size Berezinskii-Kosterlitz-Thouless

transition at grain boundaries in solid $^4$He and the role of $^3$He impurities. *Phys. Rev. Lett.* **101**, 075301 (2008).

28. Ray, M. W. & Hallock, R. B. Observation of unusual mass transport in solid hcp $^4$He. *Phys. Rev. Lett.* **100**, 235301 (2008).

29. Fear, M. J., Walmsley, P. M., Zmeev, D. E., Makinen, J. T. & Golov, A. I. No effect of steady rotation on solid $^4$He in a torsional oscillator. *J. Low Temp. Phys.* **183**, 106-112 (2016).

30. Tsuiki, T., Takahashi, D., Murakawa, S., Okuda, Y., Kono, K. & Shirahama, K. (to be published)

**METHODS**

**KAIST rigid double-torus torsional oscillator (TO).** The KAIST rigid double-torus TO is employed to reduce non-trivial viscoelastic contributions induced by the elastic property change of solid helium complicatedly coupled with the TO structure. We successfully removed three non-linear elastic contributions due to (1) uncharted shear-modulus-dependent coupling between TO parts (the glue effect)[8,31], (2) solid helium contained in the torsional rod (the torsion rod hole effect)[9], and (3) solid helium layer grown on a thin TO basal plate (Maris effect)[10]. The detailed structure of KAIST rigid double-torus TO is shown in Figure 1-(a). The most distinguishable feature compared to the non-rigid TO used for the previous rotation experiment[6,7] is that the whole structure of the TO oscillates as one rigid body. The previous non-rigid TO used several epoxy joints to connect inner and outer parts. As the temperature decreases, a gap between the inner and outer part of the TO was developed due to a thermal differential contraction. Helium leaked inside this gap and encapsulated the inner part of TO. After being solidified at a low temperature, this helium decoupled the motion of inner part from that of outer part and induced relative phase difference between them. As a result, the change in shear modulus of solid helium filled inside the gap could significantly amplified, so that it generated a large superfluid-mimicking period shift. To remove this glue effect, we assembled every joints in the TO using stainless-steel screws or hard soldering to eliminate elastic contribution on the TO response due to the relative motion between sub-parts. More details about the rigid TO construction was summarized in our previous publication[13].

Besides, the TO can be operated in two different resonant frequencies, which enables us to study a frequency-dependence of the TO response. The TO has two resonant frequencies: 448.65Hz (in-phase mode) and 1131.20Hz (out-of-phase mode). Analogous to a double-pendulum mass-spring system, two containers oscillate in same direction in the in-phase mode, while oscillate in the opposite direction in the out-of-phase mode (See Figure 1-(b)). The TO is electrically driven by ac oscillating current with a resonant frequency, $f$. Figure 1-(c) presents a schematic circuit diagram of the TO operation. The TO maintains its self-resonance through a phase-locked feedback circuit. The supplementary electrode is installed to monitor an actual motion of the upper torus. We confirmed that the phase difference between two tori was approximately 0 degrees for the in-phase mode and 180 degrees for the out-of-phase mode. The amplitude of oscillating motion was detected through a lock-in amplifier, synchronized with a reference signal provided by the main measurement scheme. The mechanical quality factor measured at 500mK is $8.58 \times 10^5$ and $4.58 \times 10^5$ in each mode. The stability of the TO period was comparable to 0.02 ns without rotation. For a 32-bar solid sample, the mass loading was calculated to 38,742 ns (5,843 ns) in the in-phase (out-of-phase) mode.

**RIKEN rotating dilution refrigerator.** To superimpose dc rotation on the torsional oscillation, the TO was attached to a mixing chamber of the rotating dilution refrigerator at RIKEN[32]. The maximum applicable rotation velocity was 4 rad/s. Since a high-Q TO technique can be sensitive to mechanical noise[33], we assessed the rotational stability (or rotational noise) of the RIKEN rotating cryostat by measuring the maximum fluctuation of rotation speed, $d\Omega$, under various rotation speed, $\Omega$. The ratio between two quantities, $d\Omega/\Omega$ was approximately 0.001 from 0 to 2 rad/s, then increased monotonically to 0.006 under 4 rad/s, comparable to that in the Manchester rotating cryostat[34]. The stability of our TO was not severely degraded by a dc rotation; it increased from 0.02 ns at a stationary state to 0.04 ns in 4 rad/s rotation velocity. To eliminate the AC torsional oscillation effect, the driving voltage of the TO was fixed at the lowest excitation voltage, 2mV (3.8mV) in the in-phase (out-of-phase) mode.

**Polycrystalline sample preparation.** We grew three polycrystalline solid helium samples with the block-capillary method. Commercially available high-purity helium with the nominal $^3$He impurity concentration of 0.3 ppm was used in the experiments. The procedure for growing solid helium was as follows: (a) we heat up the mixing chamber and the TO above a starting temperature (mostly, above 3K). (b) The sample space inside the TO was pressurized to target pressure of 60~82 bars. (c) After waiting for an equilibrium, we cooled down both mixing chamber and sample cell slowly. (d) We simultaneously monitored a change in a TO period during the cooling procedure. When a cooling curve meet a liquid-solid phase boundary, solidification was initiated. A sudden increase in the TO period is a signature for the initiation of solid growth. (e) The solidification was finished when the period of the TO reached its saturation value. Total growth time was at least longer than 2 hours for all solid samples. (f) The mixing chamber and the TO were slowly cooled down to 500mK. (g) The mass loading, a increase in the TO period due to solid mass loaded, was measured at a temperature of 500 mK.

## REFERENCES

31. Kim, D. Y. & Chan, M. H. W. Absence of supersolidity in solid helium in porous vycor glass. *Phys. Rev. Lett.* **109**, 155301 (2012).

32. Takahashi, D. & Kimitoshi, K. New rotating dilution refrigerator for a study of the free surface of superfluid He. *AIP. Conf. Proc.* **850**, 1567-1568 (2006).

33. Yagi, M., Kitamura, A., Shimizu, N., Yasuta, Y. & Kubota, M. Torsional oscillator experiments under DC rotation with reduced vibration. *J. Low Temp. Phys.* **162**, 754-761 (2011).

34. Fear, M. J., Walmsley, P. M., Chorlton, D. A., Zmeev, D. E., Gillott, S. J., Sellers, M. C., Richardson, P. P., Agrawal, H., Gatey, G. & Golov, A. I. A compact rotating dilution refrigerator. *Rev. Sci. Instrum.* **84**, 103905 (2013).

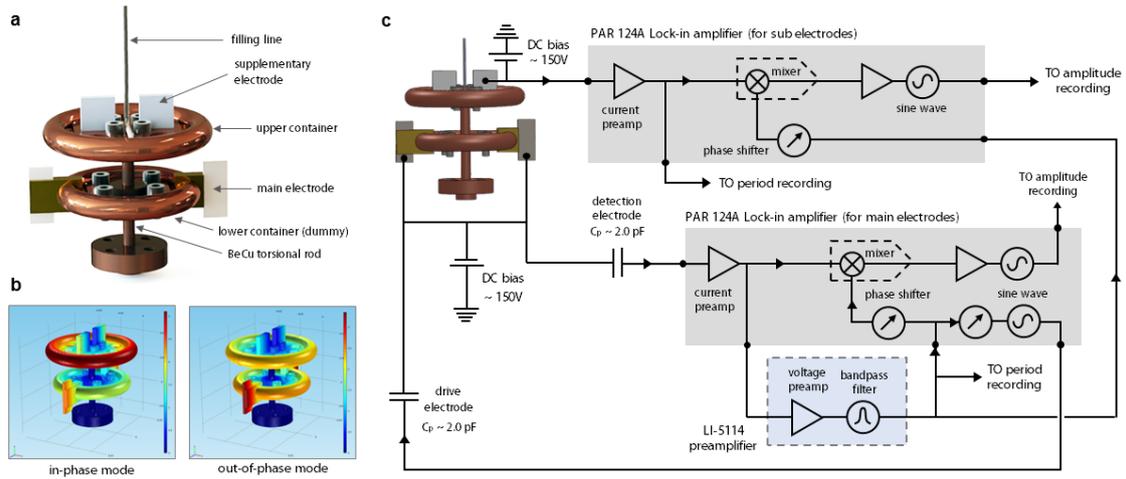

**Figure 1 | KAIST Rigid double-torus torsional oscillator (TO). a,** The two-frequency rigid TO consists of two torus-shaped helium containers and two torsional rods. A pair of supplementary and main capacitive electrodes are attached to both upper and lower container, respectively. The main electrodes are used for both drive and detection of the TO, while the supplementary electrodes are used for only detection of the actual displacement of upper torus. **b,** Similar to a double pendulum, the KAIST double-torus TO has two different resonant torsional oscillation modes. Two containers oscillate in same direction in the in-phase mode (448.65Hz), while oscillate in the opposite direction in the out-of-phase mode (1131.20Hz). Two operation modes of the TO are demonstrated through FEM simulation. The color map in the figures shows the actual displacement (red-color means high displacement). **c,** Schematic representation of a double-frequency TO measurement is shown. The TO is driven in a high-Q self-resonant mode by a phase-locked feedback circuit maintaining its resonance. The supplementary electrodes are used for only detection of the motion of upper torus. The electrical signal from the supplementary electrode is synchronized with the reference channel of the main electrode scheme.

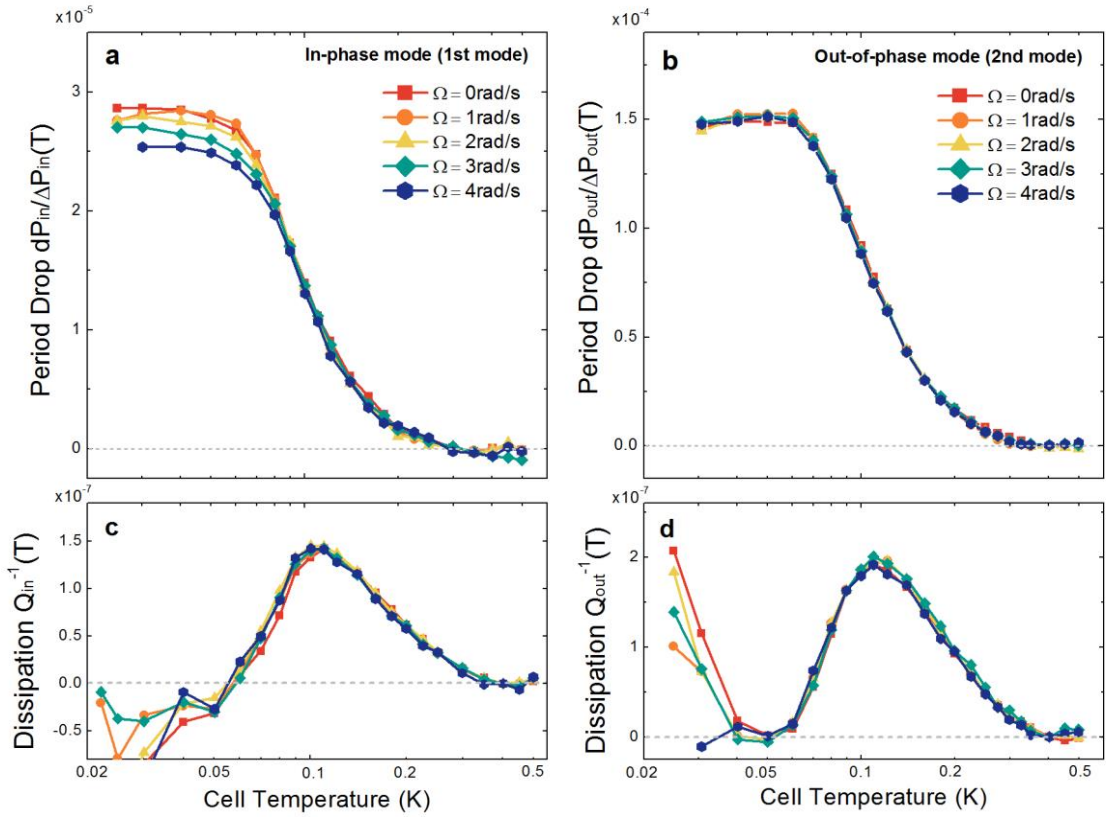

**Figure 2 | Effect of dc rotation on rigid TO responses.** Period drop $dP$ of the KAIST rigid double-torus TO, with respect to the mass loading of solid growth $\Delta P(T)$, in the **a**, in-phase and **b,** out-of-phase mode are plotted against a temperature at various dc rotation speeds in a range of 0 to 4 rad/s. Dissipation $Q^{-1}(T)$ in the **c,** in-phase and **d,** out-of-phase mode is also measured simultaneously. The same color-coded pair for each mode indicates the data obtained at same DC rotation speed. The empty-cell background was subtracted. To eliminate ac torsional oscillation effect, the driving voltage of the TO was fixed at the lowest excitation voltage, 2mV (3.8mV) in the in-phase (out-of-phase) mode.

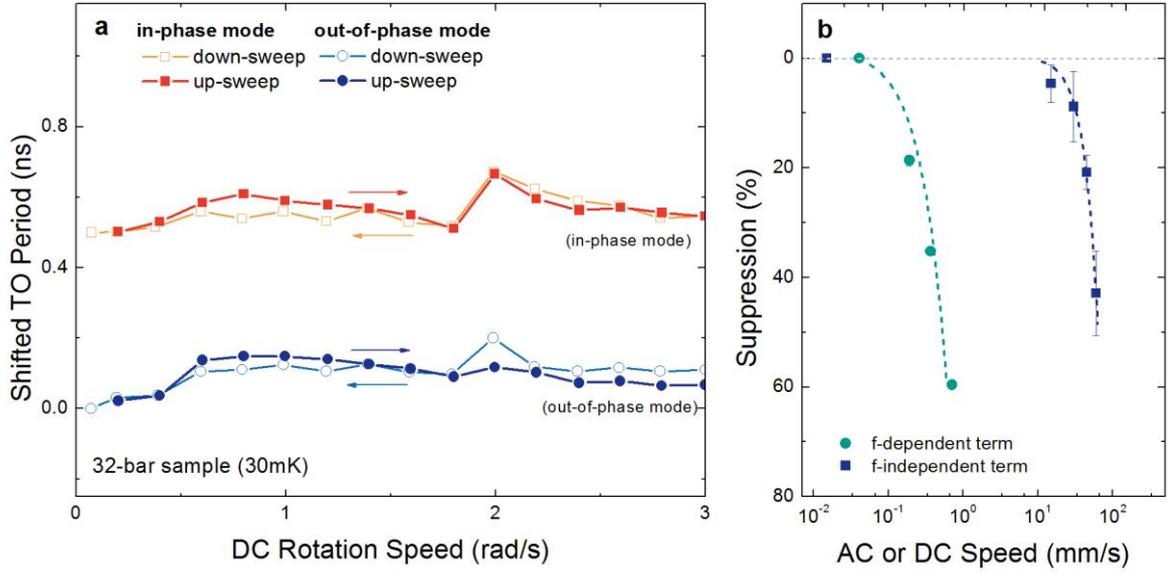

**Figure 3 | Dependence of the rigid TO responses on dc rotation or ac oscillation speed. a,** Rotation velocity-sweep scans were performed at a constant temperature. We first set the rotation speed to 3 rad/s at 500mK and then cooled the TO to a target temperature (30 mK). When the TO settled down at the target temperature, the rotation speed was swept down from 3 rad/s to 0 rad/s (open marks) with a discrete step of 0.2 rad/s, then swept back up to 3 rad/s (solid marks). Both the period and amplitude of the TO were measured for 0.5 hours at each point. The measured period values $P$ were shifted by $P_0$, for more clear representation; the value of $P_0$ is 2,267,631.549 ns (884,013.648 ns) for the in-phase (out-of-phase) mode. **b,** Suppression percentage of frequency-independent term $(dP_{in}/\Delta P_{in}(T))^{ind}$ and frequency-dependent term $(dP_{in}/\Delta P_{in}(T))^{dep}$ measured below the saturation temperature (~50mK) was plotted against a dc rotation speed or ac oscillating speed, respectively.

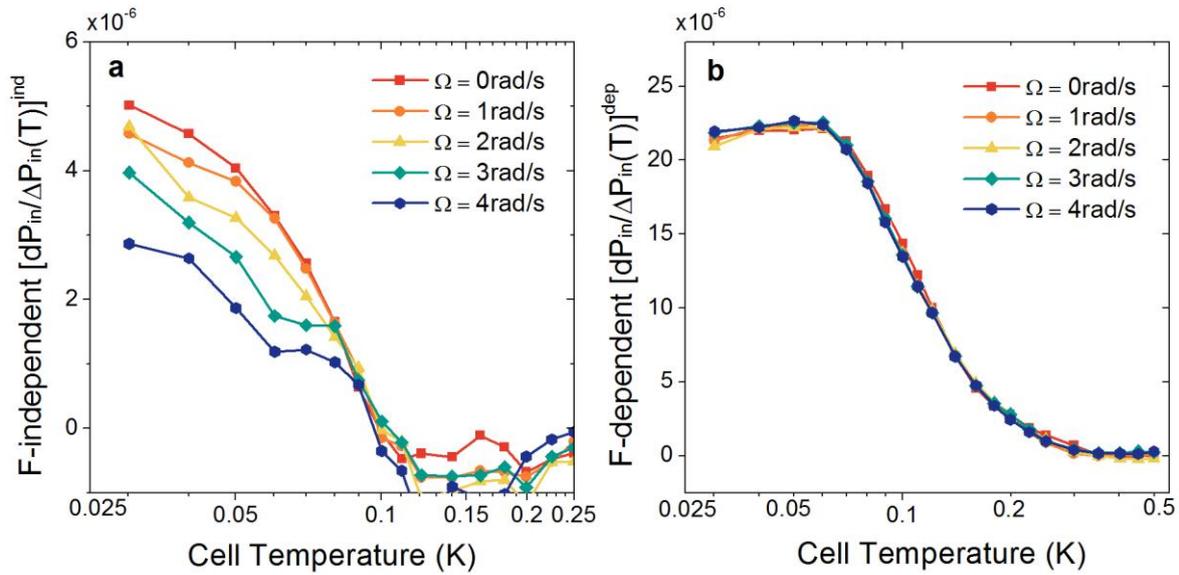

**Figure 4 | Frequency-dependent analysis of the rigid TO responses under various DC rotation speed.** Both **a,** Frequency-independent term $(dP_{in}/\Delta P_{in}(T))^{ind}$ and **b,** frequency-dependent term $(dP_{in}/\Delta P_{in}(T))^{dep}$ was decomposed explicitly at the lowest an AC oscillating speed. The same color-coded pair for each mode indicates the data obtained at same DC rotation speed.

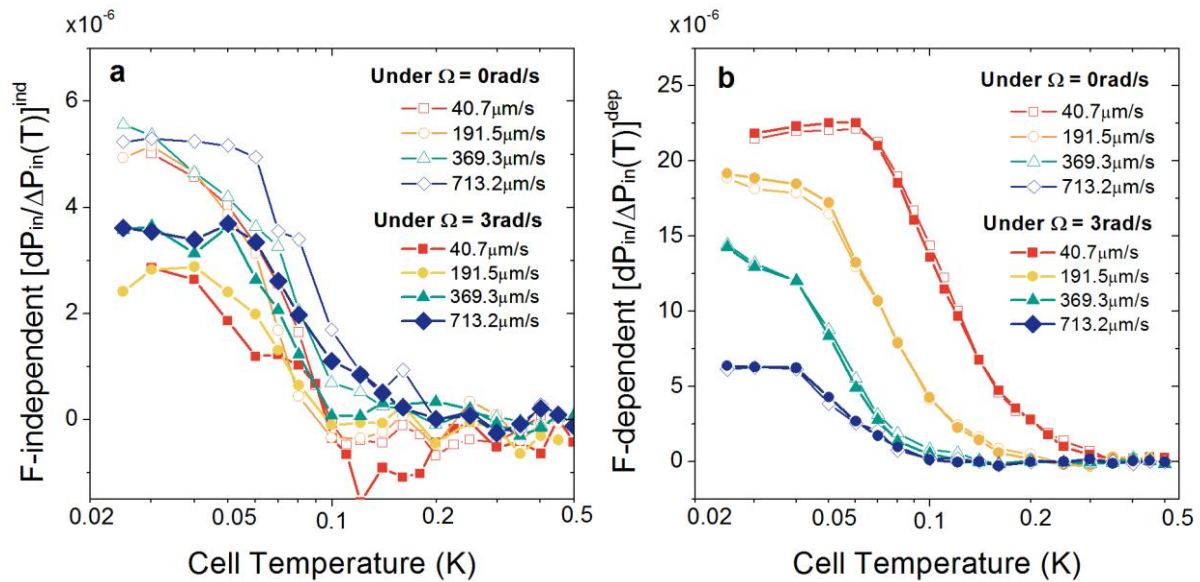

**Figure 5 | Frequency-dependent analysis of the rigid TO responses at various AC oscillating speed.** Both **a,** Frequency-independent term $(dP_{in}/\Delta P_{in}(T))^{ind}$ and **b,** frequency-dependent term $(dP_{in}/\Delta P_{in}(T))^{dep}$ was decomposed explicitly, under constant DC rotation speed of 0 (open marks) and 3 rad/s (solid marks). The same color-coded pair for each mode indicates the data obtained at same AC oscillation speed. Here, the AC driving voltage was carefully adjusted so that the solid sample experienced the same oscillating speed for the in-phase and out-of-phase mode.